\pdfoutput=1
\documentclass[%
	amsmath,
	amssymb,
	aps,
	pre,
	reprint,
	longbibliography,
	superscriptaddress,
	footinbib,
	noshowpacs,
	showkeys,
	onside,
	byrevtex,
  ]{revtex4-1}
\usepackage{ifpdf}
\usepackage{mathrsfs}[1996/01/01]
\usepackage[rgb,dvipsnames,x11names]{xcolor}
\usepackage{graphicx}
\ifpdf
\usepackage{hyperref}
\usepackage[all]{hypcap}
\usepackage{bookmark}
\fi
\usepackage{filecontents}

\DeclareGraphicsExtensions{.mps}
\ifpdf
\hypersetup{%
	pdfdisplaydoctitle=true,
	pdftitle={Structure Entropy, Self-Organization, and Power Laws in Urban Street Networks: Evidence for Alexander's Ideas},
	pdfauthor={J\'er\^ome~Benoit (ORCID: 0000-0003-1226-6757) and Saif~Eddin~Jabari (ORCID: 0000-0002-2314-5312)},
	pdfsubject={
		Subject Areas: 
			Complex Systems,
			Interdisciplinary Physics,
			Statistical Physics%
		},
	pdfkeywords={
		urban street networks;
		self-organizing systems;
		entropic equilibrium;
		MaxEnt;
		Pareto;
		power law;
		complex systems;
		statistical physics;
		information physics;
		social physics;
		partial-order;
		Galois lattice;
		surprisal;
		Christopher Alexander%
		},
	pdfcreator={REV\TeX\ 4-1 and its friends},
	pdfpagelayout=SinglePage,
	pdfpagemode=UseOutlines,
	pdfstartpage=1,
	pdfhighlight=/O,
	pdfview=FitH,
	pdfstartview=FitH,
	allcolors=RedOrange,
	colorlinks=true,
	citecolor=RoyalBlue3,
	urlcolor=RoyalBlue3,
	linkcolor=RoyalBlue3,
	bookmarksnumbered=true,
	bookmarksopen=true,
	bookmarksopenlevel=3,
	}
\fi

\begin{document}
\title[Structure Entropy, Self-Organization, and Power Laws in Urban Street Networks]{
	Structure Entropy, Self-Organization, and Power Laws\protect\\
	in Urban Street Networks:
	Evidence for Alexander's Ideas
	}
\author{\firstname{J\'er\^ome} \surname{Benoit}}%
\email[Corresponding author: ]{jerome.benoit@nyu.edu}
\affiliation{%
	New~York~University Abu~Dhabi,
	Saadiyat~Island,
	POB~129188,
	Abu~Dhabi,
	UAE%
	}
\author{\firstname{Saif Eddin} \surname{Jabari}}%
\affiliation{%
	New~York~University Abu~Dhabi,
	Saadiyat~Island,
	POB~129188,
	Abu~Dhabi,
	UAE%
	}
\affiliation{%
	New~York~University Tandon~School~of~Engineering,
	Brooklyn,
	NY 11201,
	New~York,
	USA%
	}
\date{\href{https://arxiv.org/abs/1902.07663v2}{arXiv:physics/1902.07663v2 [physics.soc-ph]}}
\begin{abstract}
Easy and intuitive navigability is of central importance in cities.
The actual scale-free networking of urban street networks in their topological space,
where navigation information is encoded by mapping roads to nodes and junctions to links between nodes,
has still no simple explanation.
Emphasizing the road-junction hierarchy in a holistic and systematic way leads us
to envisage urban street networks as evolving social systems
subject to a Boltzmann-mesoscopic entropy conservation.
This conservation,
which we may interpret in terms of surprisal,
ensures the passage from the road-junction hierarchy to a scale-free coherence.
To wit,
we recover the actual scale-free probability distribution
for natural roads in self-organized cities.
We obtain this passage
by invoking Jaynes's Maximum Entropy principle (statistical physics),
while we capitalize on modern ideas of quantification (information physics)
and well known results on structuration (lattice theory)
to measure the information network entropy.
The emerging paradigm,
which applies to systems with more intricate hierarchies as actual cities,
appears to reflect well the influential ideas on cities of the urbanist Christopher~Alexander.
\end{abstract}
\maketitle

\section{\label{sec/introduction}Introduction}

Understanding the information encoded
in the structures of complex transportation systems
is a puzzling challenge
in complex systems and statistical physics.
Information, energy, or materials
circulate
through disparate systems,
in various effective fashions,
and in multiple forms.
Vascular systems that spread blood and sap,
neural dendrites that transmit signals,
and
river basins that drain water
offer interesting examples.
Their evident diversity and complexity
mask a striking regularity along simplicity.
Their transportation structures
actually
undergo a scaling law
that reveals a simple underlying principle.
Vascular systems make their exchange surface areas \textit{``maximally fractal''} \cite{GBWestJHBrownBJEnquist1999},
dendritic trees minimize their wirings \cite{HCuntzAMathyMHausser2012},
river networks span minimally \cite{JRBanavarAMaritanARinaldo1999}.

Within this perspective,
this paper aims
to reveal the scaling principle that
drives urban street networks.
City related transportation networks
had been
for complex systems
a fruitful source of case studies
before the internet age \cite{NSBarabasi2016,MEJNewmanNI,WattsStrogatz1998,BarabasiAlbert1999,ADBroidoAClauset2018,PHolme2019,BHillier1999,PAFTurner2001,BJiangTAUSN2002,BJiangTAUSN2004}.
The electrical power grid of the western United States
provided evidence for the two instrumental breakthroughs
in the renewal of network studies \cite{WattsStrogatz1998,BarabasiAlbert1999,ADBroidoAClauset2018,PHolme2019}.
One of them is
the discovery of
\textit{scale-free} hierarchies among real-world networks \cite{BarabasiAlbert1999,ADBroidoAClauset2018,PHolme2019}.
Urban street networks appeared promptly
to undergo scale-free behaviours
as well
\cite{BJiangTAUSN2004,CrucittiCMSNUS2006,PortaTNAUSPA2006,PortaTNAUSDA2006,BJiangATPUSN2007,BJiangSZhaoJYin2008,BJiangTSUSNPDC2014}.
Keeping focus on their scaling property
and adopting approaches from the urban community
gradually
led us
to a tractable scaling model
actually
driven by a simple conservation principle.
In pursuing our goal,
we realized that our modelization was shedding a new light
on the authoritative urban theory
developed by C.~Alexander \cite{CAlexanderACINAT1965,CAlexanderTNOOSet,NASalingarosPUS}.
We will interpret our scaling model accordingly.

If the urban community has looked at urban street networks
for additional traits with tools developed for generic networks,
it has also investigated for city specific traits
with tools and approaches inherited from its own background
\cite{BHillier1999,PAFTurner2001,BJiangTAUSN2002,BJiangTAUSN2004,CrucittiCMSNUS2006,PortaTNAUSPA2006,PortaTNAUSDA2006,BJiangATPUSN2007,BJiangSZhaoJYin2008,BJiangTSUSNPDC2014,APMsucci2009,BJiangCLiu2009,MRosvall2005,APMsucci2016,CMolinero2017,BJiangACNPAW2016,CAlexanderACINAT1965,CAlexanderTNOOSet,NASalingarosPUS}.
Interestingly enough,
the insightful thought of
the urbanist
C.~Alexander on cities \cite{CAlexanderACINAT1965,CAlexanderTNOOSet,NASalingarosPUS}
has appeared to resonate with scale-free invariance
through the notion of ``natural'' city
\cite{CAlexanderACINAT1965,BJiangACNPAW2016,BJiangSZhaoJYin2008}.
``Natural'' cities refer to self-organized-like cities.
We must also mention
the more mathematically oriented
but no less insightful
work of R.~H.~Atking
on relation functions as pre-networking functions
which led to Q-analysis~\cite{RHAtkin1974,*RHAtkin1977}.

To explain the scale-freeness
of the electrical power grid,
A.-L.~Barab\'asi and R.~Albert emphasize in Ref.~\onlinecite{BarabasiAlbert1999}
two key features of real-world networks:
growth and preferential attachment.
They introduce accordingly a model
which appeared to be the \emph{Yule process} \cite{MEJNewmanPLPDZL2005,HASimon1955,*GUYule1925}:
new nodes attach to old ones with a probability proportional to the valence of the old nodes.
Mathematical analysis shows that the Yule process reproduces
the scale-freeness behaviour for nodes with high valences \cite{MEJNewmanPLPDZL2005,HASimon1955},
hence the relevance of this seminal approach \cite{BarabasiAlbert1999,MEJNewmanPLPDZL2005}.
Yule-like processes for urban street networks have been elaborated
\cite{MBarthelemy2011,MBarthelemyMUSP2008,*MBarthelemyCEDTSMCF2009,TCourtatMMCGA2011,YRui2013,APMsucci2014}.
For these adapted growth-and-preferential-attachment models,
the preferential attachment mechanism becomes local algorithms based on street-segments:
with respect to some local algorithmic policies,
the street-segments are either
budding \cite{MBarthelemyMUSP2008,TCourtatMMCGA2011,YRui2013},
fragmenting \cite{APMsucci2014},
connecting \cite{MBarthelemyMUSP2008,TCourtatMMCGA2011,YRui2013,APMsucci2014},
or a combination of them.
These models succeed to reproduce,
by varying their respective parameters,
a large variety of the patterns
effectively observed among self-organised urban street networks.
In particular,
they can reproduce scale-freeness.
However,
the intricate nature of their local algorithmic policies
renders them hardly tractable.
But still,
their success let us think that a simple Yule-like process must exist
for self-organised urban street networks.
These adaptations implicitly inject
into the Yule process
the notions of locality and globality
to which it is originally blind.
The Yule process is neither local nor global
in the sense that it involves no typical neighbourhood.
In fact,
the above adaptations mostly favour a local process over a global one:
it is essentially expected that local principles solely drive global behaviours.
While this expectation may simply misfit with a model without typical neighbourhood,
it may also lead astray by seeking finer and finer tuned local policies
that become more and more algorithmic.
Statistical physics teaches us that
such pitfalls can be addressed by
introducing a suitable global principle that
promotes pertinent traits over fine details.
Finally,
besides preferential attachment,
real-world networks
subject to scale-freeness
may also evolve
by removing, inserting, or rewiring connections \cite{BarabasiAlbert1999,MEJNewmanPLPDZL2005}.
As preferential attachment,
these connecting mechanisms involve no typical neighbourhood.
So,
they may be equally difficult to catch solely through a local policy.
In this paper,
we present
for urban street networks
a global principle that features growth, preferential attachment and favourable reattachment,
and that ultimately leads to scale-freeness.

What is original and singular
about road networks and urban street networks
is that they underlie a unique and unorthodox dual representation.
Route networks are primarily made of junctions connected by segments of routes.
On the other hand,
route networks embed roads connected by these same junctions.
The former representation
--- the \emph{primal representation} ---
corresponds to a strict \emph{geometrical representation}
of route networks as \emph{planar graphs},
while the latter
--- the \emph{dual representation} ---
interprets itself as a \emph{topological representation}
and/or \emph{information representation}
of route networks.
It is now well acknowledged that the topological space
(or information space)
captures
the complexity of route networks.
This yields the insight that
the geometry of route networks is strongly constrained by space and geography
whereas their topology reflects social, cultural, and economical activities.
For instance,
the largest number of junctions for urban street networks
have notably a valence of three or four,
while by contrast
the valence distribution of roads for self-organized urban street networks
broadly span to a scale-free power law.
The geometrical/topological duality sheds a completely new light
on the aforementioned growth-and-preferential-attachment adaptations.
These adaptations appear now to add new nodes and connections
in the \textit{``geometrically constrained space''}
even though they should rather act in the topological space.
This is not intentional.
It simply indicates that
the geometrical space is easier to apprehend than the topological space.
In this paper,
we deliberately work in the topological space.

Past results on urban description
and modern ideas of quantification and maximum entropy
render possible to approach the topological space
in a holistic and systematic fashion.
First,
we emphasize the road-junction incidence relation of urban street networks.
A \emph{natural road}
(or \emph{road})
denotes here an accepted substitute for a ``named'' street \cite{BJiangSZhaoJYin2008}.
This holistic preamble is adopted from Q-analysis \cite{RHAtkin1974}.
Subsequently Q-analysis is applied
in its paroxysmal but corrective variant
due to Y.-S.~Ho \cite{YSHoTPP1982D,SMMacgillTSpringer1987},
which is nothing but
the Formal Concept Analysis (\textsc{FCA}) paradigm \cite{BADaveyHAPriestleyILO,*GGratzerLTF}.
This paradigm builds
from the road-junction incidence relation
a one-to-one correspondence
between the topological space and a \emph{partial-order} \cite{YSHoTPP1982D,BADaveyHAPriestleyILO}.
Partial-orders are equivalent to algebraic structures,
known as \emph{Galois lattices},
which \emph{information physics} \cite{KHKnuth2011,KHKnuth2008,KHKnuth2005,KHKnuth2014}
allows to quantify and measure in a unique and systematic way
---
the involved measures being information measures.
\textit{In fine},
information physics permits us to unambiguously associate
to the topological space a functional entropy
whose the two function unknowns are meant
to describe the physics of each road or junction
and to be a probability function,
respectively.
This means that
the topological space can be interpreted as undergoing a fluctuating equilibrium
by virtue of Jaynes's \textit{Maximum Entropy principle}.
Here we envisage urban street networks as evolving social systems
subject to an entropic equilibrium
comparable to the one effectively observed
among cities of a same cultural basin \cite{YDover2004,MMilakovic2001}.
Our approach
recovers
the discrete Pareto probability distribution
(scale-free power law distribution)
widely observed
for natural roads spreading in ``natural'' cities
\cite{PortaTNAUSDA2006,BJiangATPUSN2007,BJiangSZhaoJYin2008,BJiangTSUSNPDC2014},
and foresees
a nonstandard bell-shaped distribution with a power law tail
for their joining junctions
found in agreement
with observable data
extracted from some typical ``natural'' urban street networks
(see Fig.~\ref{fig/USN/plots}).
Retrospectively,
the cohering
(or fluctuating)
part of our approach
is the Paretian match
for the Gaussian model in statistical physics,
while
the ordering
(or structuring)
part
is a reminiscence of C.~Alexander's ideas \cite{CAlexanderACINAT1965,CAlexanderTNOOSet}
(see Fig.~\ref{fig/USN/GaloisLattice}).

Although our approach is specifically applied to urban street networks,
it provides a generic paradigm for the study of complex networks
underlying partial-orders.
Within this broader perspective,
urban street networks become an ideal \emph{toy model}
and C.~Alexander's ideas fall into the domain of network theory.

The manuscript is organized as follows.
Our paradigm is sketched as the first course (Sec.~\ref{sec/paradigm}).
Then a brief survey of the state of the art in urban street networks
is given before we proceed forwards (Sec.~\ref{sec/USN}).
Once the paradigm is applied,
we discuss further our results
from the perspective of C.~Alexander's ideas (Sec.~\ref{sec/Discussion}).
Eventually,
we point to future investigations (Sec.~\ref{sec/conclusion}).

\section{\label{sec/paradigm}Paradigm}

\subsection{\label{sec/paradigm/sbm}Structure before measure}

\subsubsection{\label{sec/paradigm/sbm/structure}Structure}

\textit{`Structure before measure (but without alteration)'}
is the dominant leitmotif of the present work.
It is borrowed from Q-analysis \cite{RHAtkin1974}
but with a severe and fundamental constraint
(in parenthesis)
after a correction \cite{SMMacgillTSpringer1987,YSHoTPP1982D}
due to Y.-S.~Ho \cite{YSHoTPP1982D}:
\textit{`We should not include anything which is not given'}.
The Q-paradigm as revisited by Y.-S.~Ho \cite{YSHoTPP1982D} leads
to plain algebraic ordering structures
known as Galois lattices \cite{YSHoTPP1982D,BADaveyHAPriestleyILO}
instead to an insightful
but \textit{in fine} deficient \cite{SMMacgillTSpringer1987,YSHoTPP1982D}
simplicial geometrical interpretation \cite{RHAtkin1974}.
As partially ordered structure,
each Galois lattice is equipped with an order relation;
as algebraic structure, with a join operator.
Two elements are either comparable or not;
an element is either join-irreducible or the join of two distinct elements.

In general,
a Galois lattice organizes itself in layers
with respect to its order relation
to give rise to a Hasse diagram \cite{BADaveyHAPriestleyILO}.
For finite distributive Galois lattices \cite{BADaveyHAPriestleyILO},
which might be considered typical \cite{BADaveyHAPriestleyILO},
the join-irreducible elements constitute the smallest nontrivial elements \cite{BADaveyHAPriestleyILO}
from which the whole builds itself through the join operator,
so that they form the lowest nontrivial layer of their Hasse diagrams.
From now on,
let us imagine this layer as a network of homogeneous elements
that links each pair of them when they can join to generate a greater element.
Along this line,
each greater element itself belongs to an upper layer envisaged
as another network of homogeneous elements
arbitrarily bonded with respect to the order relation.

\subsubsection{\label{sec/paradigm/sbm/measure}Measure}

What about \textit{`measure'} ?
As answer,
let us invoke
the formal statement that arises
from the emerging theory of information physics \cite{KHKnuth2008}:
\textit{`Measuring is the quantification of ordering'}.
More precisely,
imposing natural algebraic consistency constraints
permit us not only to evaluate
(or to quantify)
Galois lattices
but also to recover and generalise contemporary information measures
(modulo
two successive
latticial exponentiations) \cite{KHKnuth2011,KHKnuth2008,KHKnuth2005,KHKnuth2014}
---
information physics is to structures
what N{\oe}ther's theorem \cite{NOETHER,*NOETHERTAVEL} is to symmetries.
For finite distributive Galois lattices \cite{BADaveyHAPriestleyILO},
the evaluation reduces to the evaluations of their join-irreducible elements,
the constraints determining the evaluations of the join-reducible elements.
Latticial exponentiations generate distributive Galois lattices.

In other words,
we have the freedom to evaluate each join-irreducible element as we wish.
Nevertheless,
while valuation functions associated to first exponentiations
are recognized as probability distributions,
further natural consistency constraints dictate
linear combinations of the Shannon and Hartley entropies \cite{JAczelBForteCTNg1974}
as valuation functions associated to second exponentiations \cite{KHKnuth2008,KHKnuth2005}.
And,
evidently,
the valuation of the initial Galois lattice is
governed by the underlying physics,
\textit{viz.},
the evaluation of each initial join-irreducible element
is meant to
express its physical state.
Meanwhile,
the probability distribution might be as plausible as possible
with respect to both
our lack of comprehensive knowledge
for each element
on their concealed microscopic details
and our macroscopic viewpoints.
This is nothing other than
Jaynes's Maximum Entropy principle
\cite{KHKnuth2008,ETJaynes1957I,*ETJaynes1957II,HKKesava2009,JNKapurHKKesava1992,ETJaynes1978SYLI}.

\subsubsection{\label{sec/paradigm/sbm/maxent}Principle of {M}aximum {E}ntropy}

Thence,
the physical content of the paradigm shifts
from an algebraic structure to a fluctuating environment,
from Galois lattice partial-order to entropic coherence.
Our initial ignorance \cite{ETJaynes1978SYLI}
yielding on the elements of the Galois lattice,
the probability distribution is
over their number of possible states.

Let $\Pr(\Omega)$ denote
the probability of an element to count $\Omega$ configurations
and
recap:
the most plausible realization of $\Pr(\Omega)$
is the one that maximizes the entropy $-\sum\Pr(\Omega)\ln\Pr(\Omega)$ \cite{FTN:EntUnits}
with suitable moment constraints known as \textit{characterizing moments} \cite{ETJaynes1957I,HKKesava2009}.
As characterizing moments,
assuming among the elements
no typical number of configurations but rather a typical scale,
we must discard any classical moment
and
may consider logarithmic moments instead.
Imposing the first logarithmic moment $\sum\Pr(\Omega)\ln\Omega$
as the unique characterizing moment
appears to lead
to the (scale-free) discrete Pareto probability distribution $\Pr(\Omega)\propto\Omega^{-\lambda}$.
Since $\ln\Omega$ measures nothing but our complete ignorance
on the state effectively occupied by any element having $\Omega$ possible states,
this constraint actually forces to preserve on average
our complete ignorance
on the elements of the Galois lattice
---
as an analogy,
the Maxwell-Boltzmann statistics describing ideal gases
can be deduced
by solely enforcing a constant mean energy \cite{ETJaynes1957I,JNKapurHKKesava1992}.

The above
\textit{deus ex machina}
has been interpreted as
some evolutionary based mechanism
to maintain some opaque internal order \cite{MMilakovic2001,YDover2004}.
Imposing the second logarithmic moment as an extra characterizing moment
leads to a statistics governed by the discrete lognormal probability distribution
$\Pr(\Omega)\propto\Omega^{-\lambda}\exp(-(\ln\Omega-\eta)^{2}/2\sigma_{\eta}^{2})$;
and so on.
For now, let us restrict ourselves to our first attempt.
In what follows,
we will denote
the cohering entropic equilibrium
governed by the discrete Pareto distribution
by \emph{Paretian coherence}.

\subsection{\label{sec/paradigm/oln}Overlying networks}

\subsubsection{\label{sec/paradigm/oln/jin}The join-irreducible network}

Now we shift our attention back
to the network formed by the join-irreducible elements of the Galois lattice.
As a working hypothesis,
let us assume for each node that
its number of configurations $\Omega$ depends on its valence~$n$;
we write $\Omega(n)$.
Therefrom,
in this network,
the probability distribution of node valences $\Pr(n)$
preserves the scale-free character
when the number of configurations $\Omega(n)$ grows powerly
according to an exponent $\nu_{1}$.
Then
we have $\Pr(n)\propto{n}^{-\lambda\nu_{1}}$
where the exponents $\lambda$ and $\nu_{1}$ characterize,
respectively,
the entropic coherence of the Galois lattice as a whole
and the configurational growth of its join-irreducible elements
as nodes of an homogeneous network.
On the other hand,
the number of configurations for every join-reducible element
remains algebraically coerced
by the Galois lattice,
that is,
it is obliged to algebraically depend on the number of configurations
of its two joining elements
through the valuation additive constraint \cite{KHKnuth2008,KHKnuth2011}.
Now,
let us envisage as a second network
the layer that gathers the joins of two join-irreducible elements,
two joins with a common generator being bonded.

\subsubsection{\label{sec/paradigm/oln/jrn}The join-reducible networks}

For the sake of argument,
let us pretend that the nodes on the first and second network-layers
undergo a powerly configurational growth
with exponents $\nu_{1}$ and $\nu_{2}$, respectively.
On our second network,
we then have
$\Pr(n)\propto\plCn(\nu_{1};n)\,n^{-\lambda\nu_{2}}$
where $\plCn(\nu_{1};n)$ counts
the occurrences of nodes of valence $n$
with respect to the valuation additive constraint,
so that it might be merely thought as a self-convolution operator
acting on the valence probability distribution of our first network.
Iterating this process gives
for the $k$-th network-layer
$\Pr(n)\propto\plCn(\nu_{1},\nu_{2},\ldots,\nu_{k};n)\,n^{-\lambda\nu_{k}}$
with obvious notations.
Thence,
under the rather favourable assumption that node configurations grow powerly with valences,
the valence probability distribution
for every reducible network-layer
inherits a power tail
from the underlying scale-free behaviour,
whereas the irreducible network-layer plainly reveals it,
and a mass function from the underlying Galois lattice algebraic structure.
Notice that
when the Galois lattice is flattened or ignored,
the valence probability distribution
sees its tail dominated by the strongest power tail
and its mass function becoming
a linear combinations of powerly weighted mass functions.

\subsubsection{\label{sec/paradigm/oln/misleadingclaim}Misleading claim}

So,
within this scheme,
we will observe no scale-free network
if the underlying ordering structure is disregarded,
if the involved network is not an irreducible one,
or if the node configurations do not grow powerly with valences.
However,
the claim that these networks are not subjected to scale-freeness
would be misleading
here
since the system as a whole is effectively driven
by a scale-free power law probability distribution,
while the scale-free behaviour could possibly be observed
only for the first network-layer.

\section{\label{sec/USN}Urban Street Networks as~Toy~Model}

\subsection{\label{sec/USN/geovtop}Geometry versus topology}

\subsubsection{\label{sec/USN/geovtop/complexities}Trivial versus nontrivial complexities}

As pedestrians, cyclists, or drivers, we tend to envision
at first glance
the junctions and street-segments of our cities as
the natural nodes and edges, respectively,
of urban street networks (see Fig.~\ref{fig/USN/networks}d).
Their complexity is nonetheless trivial:
three or four links for most street junctions \cite{BJiangSZhaoJYin2008,BJiangCLiu2009}.
Indeed,
\textit{in situ},
any city-adventurer knows that
at each street-segment-end
(or junction)
they would have in most case only two alternatives:
continue along or the other way.
This occurs independently
of the city they explore
or where they are in the city.
Unsurprisingly,
the geometrical representation
(also called primal representation)
has appeared too naive to embody the complexity of urban street networks
\cite{BJiangTAUSN2004,BJiangSZhaoJYin2008,BJiangCLiu2009,BJiangTSUSNPDC2014,PortaTNAUSPA2006,MRosvall2005,APMsucci2009}.

A second thought may lead us to realize that
we rather reason in terms of streets than of street-segments
--- and possibly in terms of junctions.
Indeed, from townsmen we expect concise directional answers shaped as follows:
\textit{``%
	To go to Oasis office from Amethyst area:
	take Sunshine street, then Seaport street
	--- at Jade junction ---
	and, finally, Sunset street
	--- at Jonquil junction.%
	''}
Even though colourful,
this typical directional answer implicitly reveals precious information:
(i)~%
neither position nor distance is expected;
(ii)~%
each junction in itself plays a secondary role;
(iii)~%
each pair of successive streets critically shares a common junction
--- whichever it is.
To wit, we expect topological responses.
The topological representation
(or dual representation)
maps streets to nodes and links each pair of them
that shares a common junction (see Fig.~\ref{fig/USN/networks}e).
In contrast to geometrical networks,
topological networks exhibit small-world and scale-free properties,
that is,
complex network behaviours~\cite{BJiangSZhaoJYin2008,BJiangCLiu2009,BJiangTSUSNPDC2014,PortaTNAUSPA2006,MRosvall2005,APMsucci2009,APMsucci2016}.
Note that topological networks can be viewed as encodings of distanceless information
which are useful for navigating through urban street networks.
For this reason the topological representation is also referred to as the information space.

\subsubsection{\label{sec/USN/geovtop/data}Data extraction overview}

Thus far
we have neglected to ask ourselves how to define streets.
This question should seem preposterous for most of us living in towns
for which a cadaster has been scrupulously maintained over decades or centuries,
but certainly not for the globetrotters among us.
Even if perfect cadasters must exist,
``named'' streets essentially remain the result of intricate social processes
where the underlying social physics likely interferes
with local customs, past or present agency struggles between social groups,
and so forth.
Actually,
the question \textit{``What is a street ?''}
has been addressed by introducing the notion of {natural road}.

\begin{figure}[bp]
	\includegraphics[width=\linewidth]{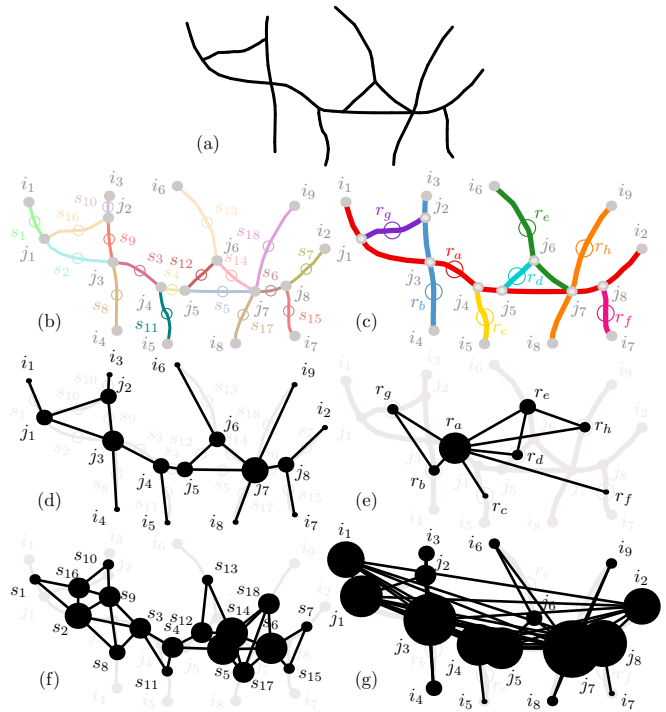}
	\caption{\label{fig/USN/networks}%
		State-of-the-art representations for urban street networks
		\cite{MRosvall2005,PortaTNAUSDA2006,BJiangSZhaoJYin2008}
		through
		a notional example \cite{FTN:NotionalExample}.
		(a)~The top row displays the mimicked \textit{`raw material'}
			as it could be extracted from any comprehensive archive.
		The left column shows the three variants of the geometrical
		(or segment-based)
		representation:
			(b)~artificially colored \textit{`raw material'} graph displaying
				street extended-junctions
				(impasses $i_{\ast}$ and effective junctions $j_{\ast}$)
				and segments $s_{\ast}$ in grey and pallid colors,
				resp.;
			(d)~junction-based connectivity graph,
				namely the concrete network without artifices;
			(f)~segment-based connectivity graph dual to graph (d).
		The right column shows the three variants of the topological
		(or natural-road-based)
		representation
		for a same natural road setup:
			(c)~revamped \textit{`raw material'} graph exhibiting
				original junctions
				and natural roads in grey and vivid colors,
				resp.;
			(e)~natural-road-based connectivity graph;
			(g)~junction-based connectivity graph dual to graph (e).
		For the four abstract networks (d-g)
		the size of each node is proportional to its valence.
		Among them,
		(e) appears as the pertinent one
		since its valence distribution
		is subject to scale-free power laws
		\cite{PortaTNAUSDA2006,BJiangATPUSN2007,BJiangSZhaoJYin2008,BJiangTSUSNPDC2014}
		(see Fig.~\ref{fig/USN/plots})
		whereas
		the ones of (d) and (f)
		are very narrow
		and the one of (g) is more intricate.
		For abstract networks (d),
		the largest number of junctions has a valence of three or four
		\cite{ACardillo2006,SLammer2006,JBuhl2006,SHYChan2011};
		so that,
		for abstract networks (f),
		the largest number of segments has a valence of four, five, or six
		\cite{BJiangSZhaoJYin2008,BJiangCLiu2009}.%
		}
\end{figure}

A natural road \cite{BJiangSZhaoJYin2008} is an exclusive sequence of successive street-segments
paired according to some \emph{behavioural based join principle} (see Fig.~\ref{fig/USN/networks}c).
Besides the \textit{de facto} cadastral join principle,
three geometrical join principles based on deflection angles \cite{BHillier1999,PAFTurner2001,MRosvall2005,BJiangSZhaoJYin2008,CMolinero2017}
are mainly used.
The \texttt{every-best-fit} join principle
is a junction-centric one
which only binds with respect to the deflection-angle-ordering of each junction,
so that it is almost deterministic
because of its local character.
The \texttt{self-best-fit} and \texttt{self[-random]-fit} join principles
are both path-centric ones
which recursively append new street-segments,
respectively,
with respect to the deflection-angle-ordering of the end-street-segments
and randomly.
Unsurprisingly,
the \texttt{self} join principles have appeared more realistic against relevant cadasters
due to their global nature
---
the \texttt{random} variant being generally the best \texttt{fit}.
Here we use the \texttt{self[-random]-fit} join principle,
unless specified otherwise.
Basically,
our \textit{`raw material'} is geometrical networks extracted from map data
fetched from well-known comprehensive archives (see Fig.~\ref{fig/USN/networks}a).

\subsection{\label{sec/USN/galoishierarchy}Galoisean hierarchy}

\subsubsection{\label{sec/USN/galoishierarchy/concealed}Concealed {G}alois lattice}

To knit a topological network we may first establish
the incidence relation $\fcaIR$
that gathers for each natural road all junctions through which it passes,
then infer its reciprocal $\fcaIR^{-1}$
that gathers for each junction all natural roads which it joins \cite{BJiangSZhaoJYin2008}:
the composition of the former with the latter $\fcaIR\mathbin{\circ}\fcaIR^{-1}$
gives the road-road topological network encountered above,
whereas the alternative composition $\fcaIR^{-1}\mathbin{\circ}\fcaIR$
leads to its dual the junction-junction topological network.
This constructive duality easily combines with the geometrical/topological duality
as exemplified in Fig.~\ref{fig/USN/networks}.
Both networks are non-injective representation of $\fcaIR$,
and so of the involved urban street network.

Let us now interpret any incidence relation $\fcaIR$ as an object/attribute relation
where each natural road acts as an object and each junction as an attribute
\cite{RHAtkin1974,BADaveyHAPriestleyILO,YSHoTPP1982D}.
Thereby,
relying on \textsc{FCA},
we can bijectively represent any incidence relation $\fcaIR$
as an ordered algebraic structure known as Galois lattice
\cite{BADaveyHAPriestleyILO,YSHoTPP1982D}.
As shown in the constructive proof provided by Y.-S.~Ho \cite{YSHoTPP1982D},
this paradigm combines objects and attributes
into pairs of subsets of them to form
without loss of information a Galois lattice.
To achieve the emerging structure,
the one-to-many relation $\fcaIR$
is naturally extended to a many-to-many relation
by stating that the attributes of two objects are their common attributes \cite{YSHoTPP1982D}.

\begin{figure}[bp]
	\includegraphics[width=\linewidth]{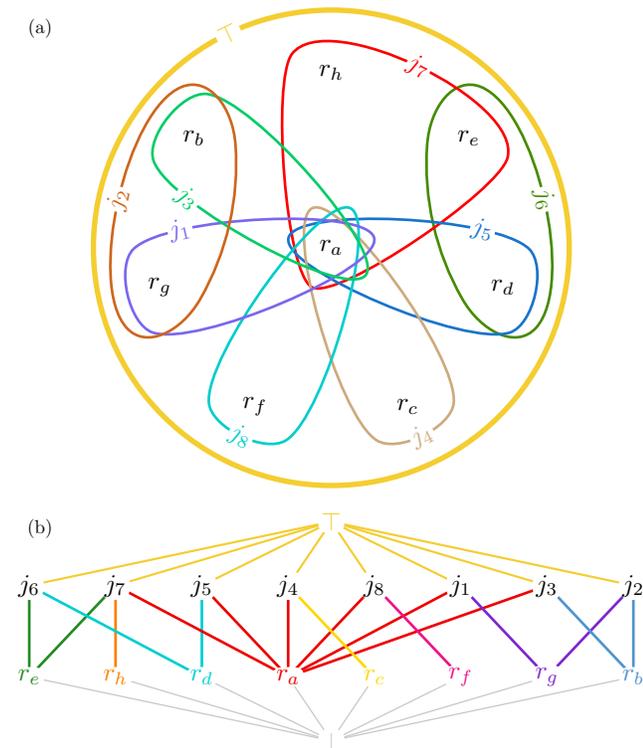}
	\caption{\label{fig/USN/GaloisLattice}%
	Illustration \textit{\`a la} C.~Alexander \cite[col.~5]{CAlexanderACINAT1965}
	for the Galois lattice
	related to the notional urban street network
	in Fig.~\ref{fig/USN/networks}.
		(a)~The subset representation
			is for evaluations of Galois lattices what a Venn diagram is for the cardinality of sets.
			The natural roads $r_{\ast}$ are singletons,
			the junctions $j_{\ast}$ are intersecting sets of natural roads,
			and the urban street network
			$\top$,
			the \emph{top element} \cite{BADaveyHAPriestleyILO},
			is the total union of the subsets.
			That is,
			natural roads $r_{\ast}$ join to form junctions $j_{\ast}$,
			while we have to be somewhere in the urban street network $\top$.
			In this work,
			the inclusion-exclusion principle for evaluations is reduced to
			its simplest nontrivial form~\eqref{eq/USN/Evaluation/constraint/addition}.
		(b)~The Hasse diagram \cite{BADaveyHAPriestleyILO} emphasizes
			the partial-order relation.
			For urban street networks, Hasse diagrams simplify in two nontrivial homogeneous layers
			---
			natural roads $r_{\ast}$ and junctions $j_{\ast}$ composing,
			resp.,
			the lower and upper layers.
			That is, natural roads $r_{\ast}$ ``pass through''
			(or imply)
			junctions $j_{\ast}$.
			The \emph{bottom element} $\bot$ is the absurd counterpart of the top element $\top$,
			{i.e.},
			emptiness.
		}
\end{figure}

Fortunately,
for urban street networks,
incidence relations
essentially maps to
Galois lattices with two nontrivial layers.
The natural roads form the lower layer
while
the junctions compose the upper one
---
the \textit{`imply'} ordering relation is ``passing through''.
In the rare event that two natural roads cross to each other more than once,
the resulting loop renders the Galois lattice more intricate.
For the sake of presentation,
we will assume that such loops are very rare or forcedly open.
Furthermore,
when every junction joins only two natural roads
the Galois lattice becomes distributive.
While mostly all junctions join only two natural roads,
we observe that
any junction that joins more than two natural roads
can be replaced by a roundabout
so that
it remains only junctions joining at most two natural roads.
For these reasons,
we will qualify as \emph{canonical}
any urban street network
whose junctions effectively join only two natural roads.
In short,
for urban street networks,
incidence relations map in a one-to-one fashion to
essentially distributive
Galois lattices
with two nontrivial layers,
while their canonicalization renders
their Galois lattices
plainly
distributive.

Arguably this is nothing new,
except that the complexity of urban street networks
can now be holistically and unambiguously measured
within the information physics framework.
The detailed treatment of this subject is
well outside the scope of this paper;
thus,
beyond the material formerly sketched (see Section~\ref{sec/paradigm}),
we simply refer to the work of K.~H.~Knuth \cite{KHKnuth2011,KHKnuth2008,KHKnuth2005,KHKnuth2014},
and we will content ourselves with presenting
the pertinent consequences for road/junction Galois lattices
to elaborate further.

\subsubsection{\label{sec/USN/galoishierarchy/complexitymeasurement}Complexity measurement}

Without loss of generality,
we may canonicalize urban street networks
so that their Galois lattices are distributive.
Henceforth,
natural roads constitute
their join-irreducible elements,
\textit{viz.},
we have the freedom to evaluate
each natural road as we desire
while the Galois lattice algebraic structure dictates to evaluate
each junction as the sum of the evaluation of their two joining natural roads.
Thusly,
for every junction $j(r,s)$ joining the pair of natural roads $(r,s)$,
we are compelled to write
\begin{equation}\label{eq/USN/Evaluation/constraint/addition}
	\glVa(j(r,s)) = \glVa(r) + \glVa(s)
\end{equation}
where $\glVa$ stands for the yet unknown valuation function.
Further consistency requirements oblige to recognize
any valuation function associated to the first exponentiation
of each Galois lattice as a probability distribution.
This probability distribution is
the composition of a yet unknown weight function $\glWg$
with the above valuation function $\glVa$;
we read
\begin{equation}\label{eq/USN/HypothesisSpace/Evaluation/composition/Pr}
	\Pr = \glWg \circ \glVa
\end{equation}
with $\Pr$ the probability distribution of the system.
Meanwhile we may choose $\glWg$ as we want.
Finally,
same and further demanded consistency constraints
force to identify the evaluation of the central element of
the second exponentiation of the Galois lattice
as the entropy $\glEta[\glVa,\glWg]$ of the system
which thusly expresses as a functional
of the valuation and weight functions,
$\glVa$ and $\glWg$,
respectively.
For canonical urban street networks,
the
functional
structure entropy $\glEta[\glVa,\glWg]$
takes the form
\begin{equation}\label{eq/USN/StructureEntropy}
	\glEta[\glVa,\glWg] =
	\!\sum_{r}\left(\glHm\circ\glWg\right)\left(\glVa(r)\right)
	+%
	\!\!\!\sum_{j(r,s)}\!\left(\glHm\circ\glWg\right)\left(\glVa(r)\!+\!\glVa(s)\right)
\end{equation}
where the first summation runs over the natural roads~$r$
and the second one over the junctions $j(r,s)$ joining the pair of natural roads $(r,s)$,
while $\glHm\colon{x}\mapsto-x\ln{x}$ is
the Shannon entropy function \cite{FTN:EntUnits}.

By reverting addition rule \eqref{eq/USN/Evaluation/constraint/addition}
in the right summation
and then composing according to \eqref{eq/USN/HypothesisSpace/Evaluation/composition/Pr},
the reader will readily recover the \textit{`flat'}
expression of the functional entropy $\glEta[\glVa,\glWg]$,
namely
\begin{math}
	\glEta[\Pr] = \sum_{e} \left(\glHm\circ\Pr\right)\left(e\right)
\end{math}
where the summation occurs indiscriminately over all natural roads and junctions~$e$.

Therefore,
in our context,
the novelty brought by information physics theory sums up as follows:
it enables us to measure the complexity of our heterogeneous system as a whole
by taking
its ordering hierarchy into account.
In detail,
it articulates as follows:
locally,
it reveals how
the natural roads $r$ impose their arbitrary valuations $\glVa(r)$ to the junctions $j$;
globally,
it unveils how
an arbitrary weight function $\glWg$ cements the whole.
Notice the slight abuse of language used
in the article's title:
entropy~\eqref{eq/USN/StructureEntropy}
is qualified with structure
to highlight this novelty.

\subsection{\label{sec/USN/paretiancoherence}Paretian coherence}

\subsubsection{\label{sec/USN/paretiancoherence/completeignorance}Assumed complete ignorance}

In any case,
from their city,
most dwellers do not perceive the underlying Galoisean hierarchy
\textit{per se}
but rather the resulting emergent Paretian coherence.
This passage from algebraic structure to organic arrangement appears
to take place in our context
as a consequence of
Jaynes's maximum entropy principle
as outlined early (see Section~\ref{sec/paradigm}).

Formally,
we assume our complete ignorance
on what phenomena drive each natural road or junction;
so that,
the most we can state is that
each one
possesses a finite number of equally likely configurations.
Thence,
the system mean entropy $\left\langle{H}\right\rangle$ writes
\begin{equation}\label{eq/USN/SystemMeanEntropy}
	\left\langle{H}\right\rangle =
		\sum_{e} \Pr(\Omega_{e})\,\ln{\Omega_{e}}
\end{equation}
whenever every natural road or junction $e$ has reached an equilibrium;
the summation happens indiscriminately over all natural roads and junctions~$e$,
$\Pr({\Omega_{e}})$ expresses the probability
for the natural road or junction $e$ to have $\Omega_{e}$ states,
and
$\ln{\Omega_{e}}$ its Boltzmann entropy.
Then,
using the same notation,
Jaynes's maximum entropy principle
invoked with the first logarithmic moment
as unique characterizing moment
literally
holds the Shannon Lagrangian
expression
\begin{widetext}\begin{equation}%
\label{eq/USN/ShannonLagrangian}
	\mathcal{L}\left(\left\{\Pr(\Omega_{e})\right\};\lambda,\nu\right) =
		-\sum_{e} \Pr(\Omega_{e})\,\ln\left(\Pr(\Omega_{e})\right)
		-\lambda \left[
				\sum_{e} \Pr(\Omega_{e})\,\ln{\Omega_{e}} - {\left\langle{H}\right\rangle}_{0}
			\right]
		-\left(\nu-1\right) \left[
				\sum_{e} \Pr(\Omega_{e}) - 1
			\right]
\end{equation}\end{widetext}
where the first and second constraints impose
the conservation of the system mean entropy
and the normalization condition satisfied by $\Pr$,
respectively,
while
${\left\langle{H}\right\rangle}_{0}$
stands for
the constant mean entropy at which the system evolves.
Resolving \eqref{eq/USN/ShannonLagrangian} readily gives
the power law distribution
\begin{equation}\label{eq/USN/CoherenceEntropy}
	\Pr(\Omega_{e}) =
		\frac{\Omega_{e}^{-\lambda}}{Z\left(\lambda\right)}
	\quad\text{with}\quad
	Z\left(\lambda\right) = \sum_{e} \Omega_{e}^{-\lambda}
\end{equation}
as \textit{Zustandssumme}.
Explicit computation of the mean entropy \eqref{eq/USN/SystemMeanEntropy}
yields
the equation of state
\begin{equation}\label{eq/USN/CoherenceEntropy/EquationOfState}
	{\left\langle{H}\right\rangle} =
		- \frac{\partial}{\partial\lambda}\ln{Z}\left(\lambda\right)
\end{equation}
whose exploitation is deferred.

In this way,
our complete ignorance helps us to discern
a Paretian coherence,
yet not plainly perceivable,
among urban street networks.

\subsubsection{\label{sec/USN/paretiancoherence/partialknowledge}Conceded partial knowledge}

In fact
we have feigned
our complete ignorance,
at least partially:
we have blithely dismissed the underlying Galoisean hierarchy
and that natural roads and junctions
are likely driven by social interactions.
It is time now to decompose
accordingly
the probability distribution \eqref{eq/USN/CoherenceEntropy}
with respect to composition \eqref{eq/USN/HypothesisSpace/Evaluation/composition/Pr}
and addition rule \eqref{eq/USN/Evaluation/constraint/addition}.

To this purpose,
it appears convenient to
adopt an agent model \cite{YDover2004,VParunakSBruencknerRSavit2004}.
Let us adapt the network of intraconnected agents model introduced in Ref.~\onlinecite{YDover2004}
for the distribution of cities in countries,
since the involved social behaviours might be similar
---
if not the same.
As agents,
we consider the inhabitants that somehow
participate to the live activity of urban street networks \cite{CAlexanderACINAT1965}:
drivers, cyclists, pedestrians,
suppliers,
institutional agents,
residents,
and so forth.
Thusly,
each natural road
(or junction)
is a hive
whose very existence
relies on the ability for each of its agents to maintain
a crucial number of intraconnections
which is
presumed crudely equal to a constant number $\plNVCR$ (or $\plNVCJ$),
called the \emph{number of vital connections} for natural roads (or junctions),
that characterizes the urban street network.
The layout of theses intraconnections is implicitly associated to
the internal order within each natural road (or junction),
while the total number of possible layouts
is simplistically considered as
its number of states.

Suppose,
for each natural road $r$,
the number of agents to be asymptotically proportional
to the number of junction $n_{r}$ through which $r$ passes
---
the ratio $A$ being constant and sufficiently large.
This hypothesis is founded upon the extensive property of natural roads.
Then the number of states $\Omega_{r}$ for every natural road $r$ yields
\begin{subequations}\label{eq/USN/AgentBasedModel/NumberOfStates}
\begin{equation}\label{eq/USN/AgentBasedModel/NumberOfStates/NaturalRoads}
	\Omega_{r}
		= \plNStR\left(n_{r}\right)
		\simeq
			\binom{\tfrac{1}{2}A\,n_{r}\left(A\,n_{r}-1\right)}{\plNVCR}
		\simeq
			\frac{A^{2\plNVCR}}{2^{\plNVCR}{\plNVCR}!}\,n_{r}^{2\plNVCR}
\end{equation}
where the generalized binomial bracket is employed.
As concerns each junction,
the involved agents are merely the agents of the two joining natural roads combined together;
hence
the same crude maneuvers give
\begin{equation}\label{eq/USN/AgentBasedModel/NumberOfStates/Junctions}
	\Omega_{j(r,s)}
		= \plNStJ\left(n_{j}=n_{r}+n_{s}\right)
		\simeq
			\frac{A^{2\plNVCJ}}{2^{\plNVCJ}{\plNVCJ}!}\,n_{j}^{2\plNVCJ}
\end{equation}
\end{subequations}
along with some abuse of notation.

Therefrom,
the valuation function $\glVa$ arises clearly as assigning
to each natural road or junction
the number of associated agents
while
the weight function $\glWg$ asymptotically counts
the number of possible vital intraconnection layouts
(modulo normalization)
in the involved
natural road or junction
then envisioned as an intranetwork.

\subsubsection{\label{sec/USN/paretiancoherence/infocascade}Cascade of information}

We can now express
the probability for natural roads and junctions
in a more specific, perceivable fashion.
Substituting \eqref{eq/USN/AgentBasedModel/NumberOfStates/NaturalRoads}
into \eqref{eq/USN/CoherenceEntropy},
we readily obtain for natural roads
\begin{subequations}\label{eq/USN/AgentBasedModel/DistributionFunction}
\begin{equation}\label{eq/USN/AgentBasedModel/DistributionFunction/NaturalRoads}
	\Pr(n_{r}) \propto n_{r}^{-2\lambda\plNVCR}
\end{equation}
which is a scale-free power law distribution.
For the junction counterpart,
inserting instead \eqref{eq/USN/AgentBasedModel/NumberOfStates/Junctions}
into \eqref{eq/USN/CoherenceEntropy},
then gathering and counting
with respect to
the precedent probability distribution \eqref{eq/USN/AgentBasedModel/DistributionFunction/NaturalRoads}
yields
\begin{equation}\label{eq/USN/AgentBasedModel/DistributionFunction/Junctions}
	\Pr(n_{j}) \propto
		\left(
			\sum_{j(r,s)} \frac{\left[n_{j}=n_{r}+n_{s}\right]}{\left({n_{r}}{n_{s}}\right)^{2\lambda\plNVCR}}%
		\right)
		\,{n_{j}}^{-2\lambda\plNVCJ}
\end{equation}
\end{subequations}
where Iverson bracket convention is used;
the summation in parentheses is simply the self-convolution of
the natural road probability distribution \eqref{eq/USN/AgentBasedModel/DistributionFunction/NaturalRoads}.
Given a natural road $r$,
its number of junctions $n_{r}$ is nothing but essentially
its degree in the involved road-road topological network:
valence distribution \eqref{eq/USN/AgentBasedModel/DistributionFunction/NaturalRoads}
has been empirically observed in self-organized
cities \cite{PortaTNAUSDA2006,BJiangATPUSN2007,BJiangSZhaoJYin2008,BJiangTSUSNPDC2014}.
The same argument dually applies for junctions:
nonetheless,
to the best of our knowledge, valence distribution \eqref{eq/USN/AgentBasedModel/DistributionFunction/Junctions}
can be neither confirmed nor refuted by the current literature.

In practical recognitions \cite{AClausetCRShaliziMEJNewman2009},
we need to assume that the number of junctions per natural road
spans from some minimal value $\plNmR\geqslant{1}$.
Then,
the normalizing constants
for probability distributions \eqref{eq/USN/AgentBasedModel/DistributionFunction}
can be effortlessly computed
in terms of natural generalizations of known (very) special functions.
While we readily have
\begin{subequations}\label{eq/USN/AgentBasedModel/CanonicalDistributionFunction}
\begin{equation}\label{eq/USN/AgentBasedModel/CanonicalDistributionFunction/NaturalRoads}
	\Pr(n_{r}) =
		\frac{n_{r}^{-2\lambda\plNVCR}}{\sfhzeta\left(2\lambda\plNVCR;\plNmR\right)}
\end{equation}
where
\begin{math}
	\sfhzeta\left(\alpha;\plNm\right) =
		\sum_{n=\plNm}^{\infty} {n}^{-\alpha}
\end{math}
is the generalized (or Hurwitz) zeta function \cite{AClausetCRShaliziMEJNewman2009,HBMF},
we find that
\begin{equation}\label{eq/USN/AgentBasedModel/CanonicalDistributionFunction/Junctions}
	\Pr(n_{j}) =
		\frac{
			\sum_{n=\plNmR}^{n_{j}-\plNmR}
				\left[{n} \left(n_{j}-n\right)\right]^{-2\lambda\plNVCR}
		\,{n_{j}}^{-2\lambda\plNVCJ}
		}{
		\sfWitten\left(2\lambda\plNVCR,2\lambda\plNVCR,2\lambda\plNVCJ;\plNmR\right)%
		}
\end{equation}
\end{subequations}
where
\begin{math}
	\sfWitten\left(\alpha,\beta,\gamma;\plNm\right) =
		\sum_{{m},{n}\geqslant\plNm}
			{m}^{-\alpha} {n}^{-\beta} \left(m+n\right)^{-\gamma}
\end{math}
is the
two-dimensional
generalized
(or Hurwitz-)
Mordell-Tornheim-Witten zeta function \cite{JMBorweinKDilcher2018}.
The former probability distribution \eqref{eq/USN/AgentBasedModel/CanonicalDistributionFunction/NaturalRoads}
is known as the \textit{discrete Pareto distribution}
and is a shifted
(or Hurwitz)
version of the better known \textit{Zipf distribution} \cite{MEJNewmanPLPDZL2005,AClausetCRShaliziMEJNewman2009};
the latter \eqref{eq/USN/AgentBasedModel/CanonicalDistributionFunction/Junctions}
is a nonstandard bell-shaped distribution
with a power law tail asymptotic to ${n_{j}}^{-2\lambda\left(\plNVCR+\plNVCJ\right)}$,
as far as we can tell,
and
we have found it convenient to name it
the \textit{Schwitten distribution}~\cite{FTN:Schwitten}.

\begin{figure*}
	\includegraphics[width=\linewidth]{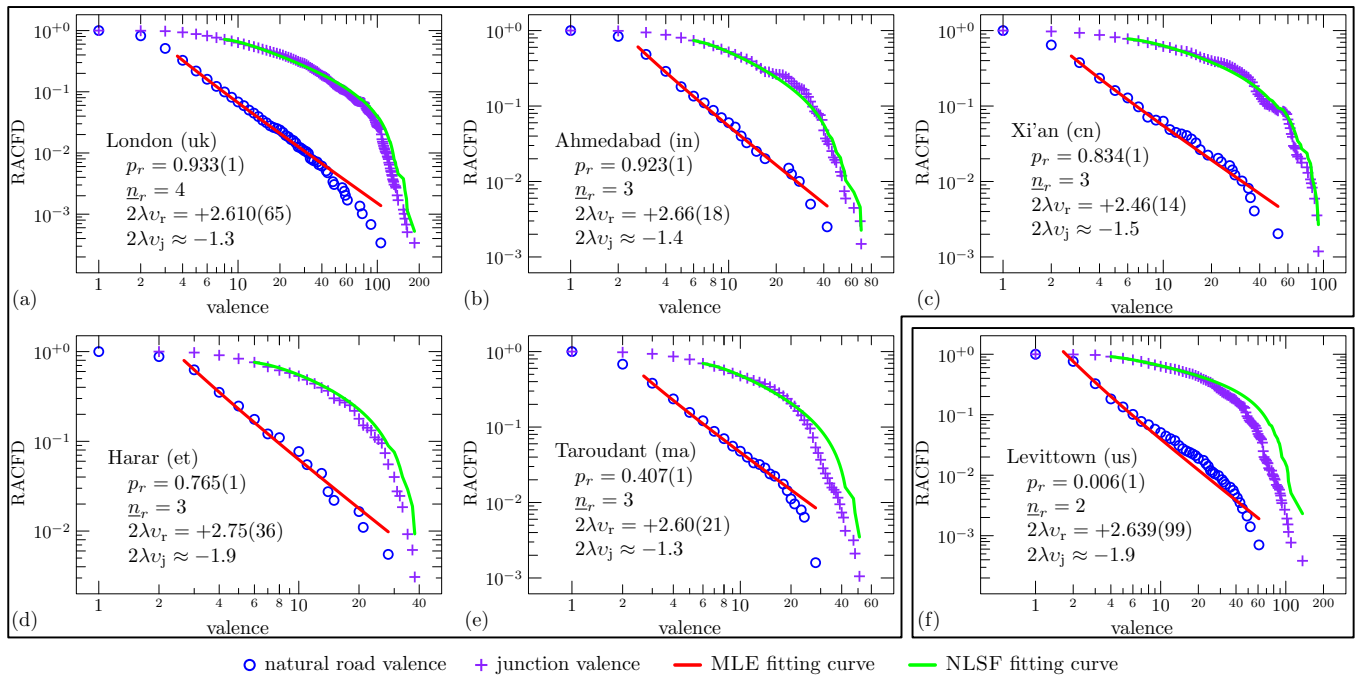}
	\caption{\label{fig/USN/plots}%
		Relative Anti-Cumulative Frequency Distributions (\textsc{RACFD})
		for five ``natural'' urban street networks (a-e) of cities with distinct cultural backgrounds
		and for an ``artificial'' urban street network (f) of a planned city:
		circles represent
		relative anti-cumulative frequencies
		for the valences of their respective road-road topological networks (see Fig.~\ref{fig/USN/networks}e);
		crosses represent
		relative anti-cumulative frequencies
		for the valences of their respective junction-junction topological networks (see Fig.~\ref{fig/USN/networks}g).
		The red fitted curves for the natural road statistics
		describe the Maximum Likelihood Estimates (\textsc{MLE})
		for the discrete Pareto probability distribution \eqref{eq/USN/AgentBasedModel/CanonicalDistributionFunction/NaturalRoads}
		estimated according to the state of the art \cite{AClausetCRShaliziMEJNewman2009,CSGillespie2015}
		($250\,000$ samples)%
		.
		The green fitted curves for the junction statistics
		show the Nonlinear Least-Squares Fittings (\textsc{NLSF})
		for the nonstandard bell-shaped discrete probability distribution \eqref{eq/USN/AgentBasedModel/CanonicalDistributionFunction/Junctions}
		with $\plNmR$ and $2\lambda\plNVCR$ fixed to
		their respective \textsc{MLE} values;
		no \textsc{MLE} approach can be computationally envisaged for the time being.
		The \textsc{MLE} goodness-of-fit qualifier $\plpVR$
		allows us to qualify
		the urban street networks
		as ``natural'' when it is greater than $0.1$,
		otherwise as ``artificial''
		\cite{AClausetCRShaliziMEJNewman2009,BJiangATPUSN2007,PortaTNAUSDA2006,PortaTNAUSPA2006,BJiangSZhaoJYin2008};
		therefore,
		our choice of urban street networks is justified \textit{a posteriori}.
		On the other hand,
		for now,
		the {ad hoc} \textsc{NLSF} data analysis prevents us from grossly rejecting
		the foreseen junction valence distribution \eqref{eq/USN/AgentBasedModel/CanonicalDistributionFunction/Junctions}.
		}
\end{figure*}

\subsection{\label{sec/USN/casestudies}Case studies}

We checked the statistical pertinence
of the foreseen junction valence distribution \eqref{eq/USN/AgentBasedModel/CanonicalDistributionFunction/Junctions}
for five urban street networks
for which
the predicted natural road valence distribution \eqref{eq/USN/AgentBasedModel/CanonicalDistributionFunction/NaturalRoads}
is a plausible hypothesis
with respect to the state-of-the-art statistical method
for power law distributions \cite{AClausetCRShaliziMEJNewman2009}
which is based on Maximum Likelihood Estimations (\textsc{MLE}).
A sixth urban street network which is recognized as planned
was taken as counter-case study.
A validation of the junction valence distribution \eqref{eq/USN/AgentBasedModel/CanonicalDistributionFunction/Junctions}
along the lines of the state of the art \cite{AClausetCRShaliziMEJNewman2009}
could not be managed
because fast evaluation of the normalizing function $\sfWitten$ has yet to be found;
meanwhile a crude data analysis based on Nonlinear Least-Squares Fittings (\textsc{NLSF})
was performed.

Figure~\ref{fig/USN/plots} exhibits
the Relative Anti-Cumulative Frequency Distributions (\textsc{RACFD})
for the valence of the road-road and junction-junction topological networks
of the six urban street networks
along with  goodness-of-fit quantifiers (or $p\text{-values}$),
the estimated parameters,
and the fitting probability distributions.
Note that the goodness-of-fit quantifiers are estimated against
the predicted natural road valence distribution \eqref{eq/USN/AgentBasedModel/CanonicalDistributionFunction/NaturalRoads}.
The \textit{`raw material'}
(see Fig.~\ref{fig/USN/networks}a)
was extracted from the Open Street Map (\textsc{OSM}) comprehensive archive \cite{OpenStreetMapContrib}.
The cities were chosen
to have distinct cultural backgrounds
and
to feature an identifiable unremodeled historical urban street network;
we picked:
(a)~%
London (United Kingdom),
(b)~%
Ahmedabad (India),
(c)~%
Xi'an (China),
(d)~%
Harar (Ethiopia),
(e)~%
Taroudant (Morocco),
and
(f)~%
Levittown (Pennsylvania, United States).
The boundary is either the innermost ring road (London),
the city wall (Ahmedabad, Xi'an, Harar, Taroudant),
or a consistent encircling series of connected roads (Levittown).
The natural roads
(see Fig.~\ref{fig/USN/networks}c)
were joined with respect to the \texttt{self[-random]-fit} join principle \cite{BJiangSZhaoJYin2008}.
For each skeleton,
we generated one hundred natural road setups,
and then we selected,
among the setups with a relatively smooth \textsc{RACFD}
for the valence of their junction-junction topological network,
the one with the highest goodness-of-fit quantifier.
Observed that for the first five urban street networks
(a-e)
the predicted natural road valence distribution \eqref{eq/USN/AgentBasedModel/CanonicalDistributionFunction/NaturalRoads}
is effectively a plausible hypothesis,
since their goodness-of-fit quantifiers $\plpVR$ are greater than $0.1$,
while for the sixth one
(f)
it must be clearly rejected \cite{AClausetCRShaliziMEJNewman2009}.
So,
as expected,
the first five are ``natural'' while the sixth is ``artificial''.

Our {ad hoc} crude data analysis appears promising
in the sense that it forbids one from grossly rebutting
the foreseen junction valence distribution \eqref{eq/USN/AgentBasedModel/CanonicalDistributionFunction/Junctions}.
Interestingly,
the case studies reveal that the number of vital connections $\plNVCJ$ is negative,
to wit
that the associated generalized binomial combination number is smaller than one.
We interpret this result as follows:
the number of agent intraconnections for junctions
is relatively much smaller than the one for natural roads.

\section{\label{sec/Discussion}Alexander's Ideas as Guide}

\subsection{\label{sec/Discussion/retocap}Retro-recapitulation}

In summary,
we can take for granted that our partial ignorance
permits us
to recognize a hierarchical Paretian coherence among urban street networks.
More precisely,
within the framework of information physics \cite{KHKnuth2011,KHKnuth2008,KHKnuth2005,KHKnuth2014},
the emerging Paretian coherence
that characterizes self-organized (or ``natural'') urban street networks
\cite{BJiangATPUSN2007,PortaTNAUSDA2006,PortaTNAUSPA2006,BJiangSZhaoJYin2008}
has not only been predicted but also shown to reveal the underlying Galoisean hierarchy
that describes any of them, either planned or self-organized.
The passage
to the Paretian coherence
---
organic by nature
---
from the Galoisean hierarchy
---
in essence algorithmic
---
occurs by imposing a logarithmic maximum-entropy constraint
with complete ignorance as the initial knowledge condition
\cite{ETJaynes1957I,HKKesava2009,JNKapurHKKesava1992,ETJaynes1978SYLI}.

Our partial knowledge hangs on
the ``passing through'' partial-ordering
that ties natural roads with junctions
and
on
the ``pairing'' that typifies any social system.
The former bijectively transforms urban street networks
into Galois lattices
whose algebraic structure,
in turn,
leads
(modulo some natural algebraic constraints \cite{KHKnuth2011,KHKnuth2008,KHKnuth2005,KHKnuth2014})
to a set of functional relations and equations
meant to measure complexity;
the latter furnishes a hint
to figure out
the two involved functional unknowns,
namely
the weight and the evaluation functions.

In the words of C.~Alexander \cite{CAlexanderACINAT1965,BJiangACNPAW2016,CAlexanderTNOOSet},
the pre-passage part is ``mechanical'';
we have used Galoisean instead.
The Formal Concept Analysis
(\textsc{FCA})
algorithmic transformation
\cite{YSHoTPP1982D,BADaveyHAPriestleyILO}
is simply a prerequisite to apply information physics \cite{KHKnuth2011,KHKnuth2008,KHKnuth2005,KHKnuth2014}.
The hint was translated to a crude asymptotic binomial paired-agent model,
which is compatible with the social machinery
taking place
``in Berkeley at the corner of Hearst and Euclid''
in Ref.~\onlinecite{CAlexanderACINAT1965}.

\subsection{\label{sec/Discussion/alexanderconjecture}Alexander's conjecture}

Convinced that nature does not like trees,
C.~Alexander
informally introduced the notion of ``semilattice'' \cite{CAlexanderACINAT1965}:
whoever has seen their hand-representations is stuck
by the resemblance between their line renderings and Hasse diagrams
before they realize that the round ones swimmingly illustrate
addition rule \eqref{eq/USN/Evaluation/constraint/addition}
(see Fig.~\ref{fig/USN/GaloisLattice}).
We believe that he
intuitively grasped the idea of the partial-ordering relation reduction to Galois lattices
---
plainly apprehended
and rigorously established
earlier by {\O}.~Ore \cite{OOre1942,OOre1944,*OOre1944:erratum}
---
along the concomitant algebraic structure~\cite{FTN:SemiLattice}.

Even so C.~Alexander did not attempt to put numbers on ``semilattices'',
he nonetheless claimed that
for ``natural'' cities
their elements
holistically arrange according to a ``living'' coherence:
it is his legacy
as urban architect.
In the literature,
it takes the form of straight lines on log-log plots
of the natural road valence distribution;
here,
for urban street networks,
it has been shown
to emerge from
Jaynes's maximum entropy principle
invoked
with the first logarithmic moment as sole characterizing moment.
Thus,
in this work,
we have
established the
statistical physics foundation
for the ``living'' coherence
occurring among ``natural'' cities,
at least for their urban street networks;
instead of ``living'' we have used Paretian.

Adopting,
as C.~Alexander might have done,
the more intuitive approach
that interprets entropy as the average amount of \emph{surprisal} \cite{MTribusTT},
the Alexander's conjecture becomes:
``natural'' cities evolve by maintaining their amount of surprisal
constant on average.
This conjecture applies to cities as a whole,
from habitations to transportation.

\subsection{\label{sec/Discussion/surprisal}Surprise}

Besides giving an intuitive macroscopic physical content,
stating Alexander's conjecture in terms of surprisal
implicitly gives to C.~Alexander's ideas a microscopic physical content.
Surprisal
(or \emph{surprise})
$\stSu=-\ln\circ\Pr$
was introduced by M.~Tribus
as a measure that quantifies our astonishment and indecision
when we face an arbitrary event \cite{FTN:EntUnits,MTribusTT}.
Along this line,
Alexander's conjecture expresses
nothing but
the conservation on average of the astonishment and indecision of dwellers
when they perceive their own city.
To draw an analogy from statistical physics,
particles of an ideal gas conserve on average their motion,
which is quantified in terms of linear momentum \cite{ETJaynes1957I,JNKapurHKKesava1992}.
So,
from a statistical physics perspective,
astonishment and indecision of dwellers of an Alexander city
appears then to be for natural roads and junctions
---
and any other similar urban items
---
what
motion is for particles of an ideal gas.

Carrying on the analogy between our system and an ideal gas
as
a parallel between a Paretian system and a Gaussian system
is relevant as well.
The distribution of number of states
would be a discrete Gaussian distribution
instead of a discrete Pareto distribution,
for the elements of the Galois lattice,
if Jaynes's maximum entropy principle
was invoked
with the first and second moments
rather than
with the first logarithmic moment
as characteristic moments.
Then the nature of the underlying discrete Gaussian distribution
might be almost preserved
for both the natural road and the junction distributions
provided that the numbers of vital connections are both equal to ${1}/{2}$.
We used the fact that the convolution of two discrete Gaussian distributions
is almost a discrete Gaussian distribution \cite{PJSzablowski2001}.
The noteworthy point is that
the junction valence distribution would then appear similar to the natural road valence distribution.
That is,
a Gaussian physics would mainly dissolve the underlying Galois lattice of our system,
while the Paretian physics presented
in this paper
reveals it.

In brief,
we are facing a Galoisean Paretian
statistical physics
that goes beyond our conventional Gaussian way of thinking \cite{BJiangACNPAW2016,FSattin2003};
C.~Alexander might have used ``mechanical'' instead \cite{CAlexanderACINAT1965,BJiangACNPAW2016,CAlexanderTNOOSet}.

\section{\label{sec/conclusion}Conclusion}

We have investigated scale-free networking in urban street networks.
Natural-road-based connectivity graphs
have been widely observed to
realize scale-free networks
in self-organized cities
\cite{PortaTNAUSDA2006,BJiangATPUSN2007,BJiangSZhaoJYin2008,BJiangTSUSNPDC2014}
---
a natural road
(or road)
is an accepted substitute for a ``named'' street \cite{BJiangSZhaoJYin2008}.
Our approach emphasizes in a holistic and systematic way
the road-junction hierarchy of urban street networks.
This approach leads to a one-to-one correspondence
between urban street networks and algebraic structures known Galois lattices,
so that
it fits with the mindset of information physics \cite{KHKnuth2011,KHKnuth2008,KHKnuth2005,KHKnuth2014}.
Ultimately,
this switch to a different framework
allows us
to envisage urban street networks as evolving social systems
subject to an entropic equilibrium \cite{MMilakovic2001,YDover2004}.
We have shown that
the passage from the underlying Galoisean
(or road-junction)
hierarchy to an underlying Paretian
(or scale-free)
coherence
can be achieved by invoking Jaynes's Maximum Entropy principle
with the first logarithmic moment as the sole characterizing constraint
and our complete ignorance as initial knowledge
\cite{ETJaynes1957I,HKKesava2009,JNKapurHKKesava1992,ETJaynes1978SYLI,MMilakovic2001,YDover2004}.
Eventually the underlying Paretian coherence
must be decomposed
with respect to the underlying Galoisean hierarchy
within the framework of information physics.
Our decomposition envisions natural roads and junctions
as hives of social agents \cite{YDover2004,VParunakSBruencknerRSavit2004}.
Social interactions are typified
by a binomial paired-agent model taken at the asymptotic limit \cite{YDover2004}.
We have recovered the discrete Pareto probability distribution
widely observed for natural roads evolving in self-organized cities
\cite{PortaTNAUSDA2006,BJiangATPUSN2007,BJiangSZhaoJYin2008,BJiangTSUSNPDC2014}.
What is more interesting,
however,
is that we have also been able to foresee
a nonstandard bell-shaped distribution with a power law tail
for their junctions.

Beyond urban street networks,
we have argued that our paradigm reflects C.~Alexander's ideas
on cities \cite{CAlexanderACINAT1965,CAlexanderTNOOSet}.
From the viewpoint of statistical physics,
the passage from Galoisean hierarchy to Paretian coherence
looks like a missing piece of his ideas.
This passage has given place to
a concise eponymous conjecture
expressed in terms of surprisal \cite{MTribusTT}.
Surprisal quantifies the astonishment and indecision of city-dwellers,
which are for Paretian statistical physics of ``natural'' cities
what motion is for Gaussian statistical physics of
ideal gases \cite{ETJaynes1957I,JNKapurHKKesava1992}.
Ultimately
we are facing
a Galoisean Paretian statistical physics that challenges
our ``mechanical'' and Gaussian ways of thinking
\cite{BJiangACNPAW2016,CAlexanderACINAT1965,CAlexanderTNOOSet,FSattin2003}.

We have also shed a new light on
how power law phenomena can emerge
from complex systems that underlie a Galoisean hierarchy.
Here
urban street networks
constitute an ideal toy model
as they mainly reduce to intuitive two-layer Galois lattices.
In this regard
we believe that scale-free networks are omnipresent
in nature
but also
that
neither their underlying partial-order
nor the logarithmic character of their statistics
have been plainly taken into account.


\begin{thebibliography}{77}
\bibitem [{Note1()}]{FTN:EntUnits}%
  \BibitemOpen
  \bibinfo {note} {Entropies and surprisals are expressed in
  \protect\textsf{nat} units.}\BibitemShut {Stop}%
\end{thebibliography}
%

\end{document}